\documentclass[letterpaper, 10 pt, conference]{ieeeconf}  
\IEEEoverridecommandlockouts
\usepackage{cite}
\usepackage{amsmath,amssymb,amsfonts}
\usepackage{algorithmic}
\usepackage{graphicx}
\usepackage{textcomp}
\usepackage{xcolor}
\def\BibTeX{{\rm B\kern-.05em{\sc i\kern-.025em b}\kern-.08em
    T\kern-.1667em\lower.7ex\hbox{E}\kern-.125emX}}
    
\usepackage{xcolor}

\newtheorem{assumption}{Assumption}
\usepackage{algorithm}
\usepackage{array}
\newcolumntype{C}[1]{>{\centering\let\newline\\\arraybackslash\hspace{0pt}}m{#1}}
\newcolumntype{L}[1]{>{\raggedright\let\newline\\\arraybackslash\hspace{0pt}}m{#1}}
\usepackage{siunitx}
\usepackage{threeparttable}
\usepackage{multirow}
\usepackage{makecell}
\usepackage{booktabs}

\usepackage{hyperref} 
\hypersetup{hidelinks,
	colorlinks=true,
	allcolors=black,
	pdfstartview=Fit,
	breaklinks=true
}
	
\begin{document}

\title{ \LARGE \bf
Motion Magnification in Robotic Sonography: \\ 
Enabling Pulsation-Aware Artery Segmentation 
}

\author{Dianye Huang$^{1, \ast}$, Yuan Bi$^{1, \ast}$, Nassir Navab$^{1, 2}$, \IEEEmembership{Fellow, IEEE}, Zhongliang Jiang$^{1, \dag}$ 
\thanks{$\ast$ Contribute equally. $\dag$ Corresponding author (e-mail: zl.jiang@tum.de).}
\thanks{$^{1}$ Dianye Huang, Yuan Bi, Nassir Navab and Zhongliang Jiang are with the Chair for Computer Aided Medical Procedures
and Augmented Reality (CAMP), Technical University of Munich, 85748
Garching, Germany.}
\thanks{$^{2}$ Nassir Navab is also with the Laboratory for Computational Sensing and Robotics, Johns
Hopkins University, Baltimore, MD 21218 USA.}
}
\maketitle

\begin{abstract}
Ultrasound (US) imaging is widely used for diagnosing and monitoring arterial diseases, mainly due to the advantages of being non-invasive, radiation-free, and real-time. In order to provide additional information to assist clinicians in diagnosis, the tubular structures are often segmented from US images. To improve the artery segmentation accuracy and stability during scans, this work presents a novel pulsation-assisted segmentation neural network (PAS-NN) by explicitly taking advantage of the cardiac-induced motions. Motion magnification techniques are employed to amplify the subtle motion within the frequency band of interest to extract the pulsation signals from sequential US images. The extracted real-time pulsation information can help to locate the arteries on cross-section US images; therefore, we explicitly integrated the pulsation into the proposed PAS-NN as attention guidance. Notably, a robotic arm is necessary to provide stable movement during US imaging since magnifying the target motions from the US images captured along a scan path is not manually feasible due to the hand tremor.
To validate the proposed robotic US system for imaging arteries, experiments are carried out on volunteers' carotid and radial arteries. The results demonstrated that the PAS-NN could achieve comparable results as state-of-the-art on carotid and can effectively improve the segmentation performance for small vessels (radial artery).
\end{abstract}


\section{Introduction}
Peripheral artery disease (PAD) is among the most impactful cardiovascular conditions \cite{smolderen2022advancing}. It is a chronic vascular disease where the arteries become blocked or narrowed, reducing the blood flow. Regular ultrasound (US) examination can effectively diagnose PAD, allowing for the early treatment. 
Considering the limited medical workforce and highly operator-dependent nature of US scans, developing a robotic US system (RUSS) for the daily US examination is becoming a research hot spot. Extensive studies have been conducted on the tasks of autonomous arteries screening, such as carotid ~\cite{de2020automated} and radial arteries~\cite{Jiang2022RAL} \textit{etc.}. To provide valuable information for further diagnosis, it is crucial to robustly and accurately extract the tubular structures from US images.  

To this end, active contour algorithm and its variants are proposed for US 3D vessel reconstruction since last decades, where the boundary is determined by minimizing predefined energy functions derived from the prior knowledge of the shape and user-determined anchor points~\cite{ukwatta2011three}. 
However, these methods require human interactions for initialization. 
Alternatively, learning-based approaches enable end-to-end fully autonomous segmentation process
and therefore start becoming mainstream after Ronneberger~\textit{et al.} proposed U-Net for biomedical image processing \cite{ronneberger2015u}. To further improve the segmentation result, attention mechanism, such as attention gated models \cite{schlemper2019attention}, and Global and Local Feature Reconstruction modules \cite{song2022global}, is incorporated into U-Net to force the network to focus more on relevant regions of the input images. Besides, US artery segmentation networks can also take advantage of US modality and the spatial continuity of the artery between two consecutive frames. For example, Z. Jiang~\textit{et al.} induced an optical flow image to increase the segmentation accuracy of limb artery \cite{jiang2022towards}; B. Jiang~\textit{et al.} optimized the femoral and tibial artery segmentation using color Doppler US images \cite{jiang2021automatic}. However, the Doppler signal is sensitive to the contact condition and the angle between the US waves and contact surface. 
Inspired by this observation, we consider that the natural pulsation of arteries caused by the cardiac-induced pressure wave \cite{alastruey2012arterial} could be employed as a stable biomedical guidance assisting the vessel segmentation. Additionally, the pulsation motion can be further used for distinguishing arteries and veins.  
\begin{figure}[tb]
     \centering
     \includegraphics[width=0.46\textwidth, angle=0]{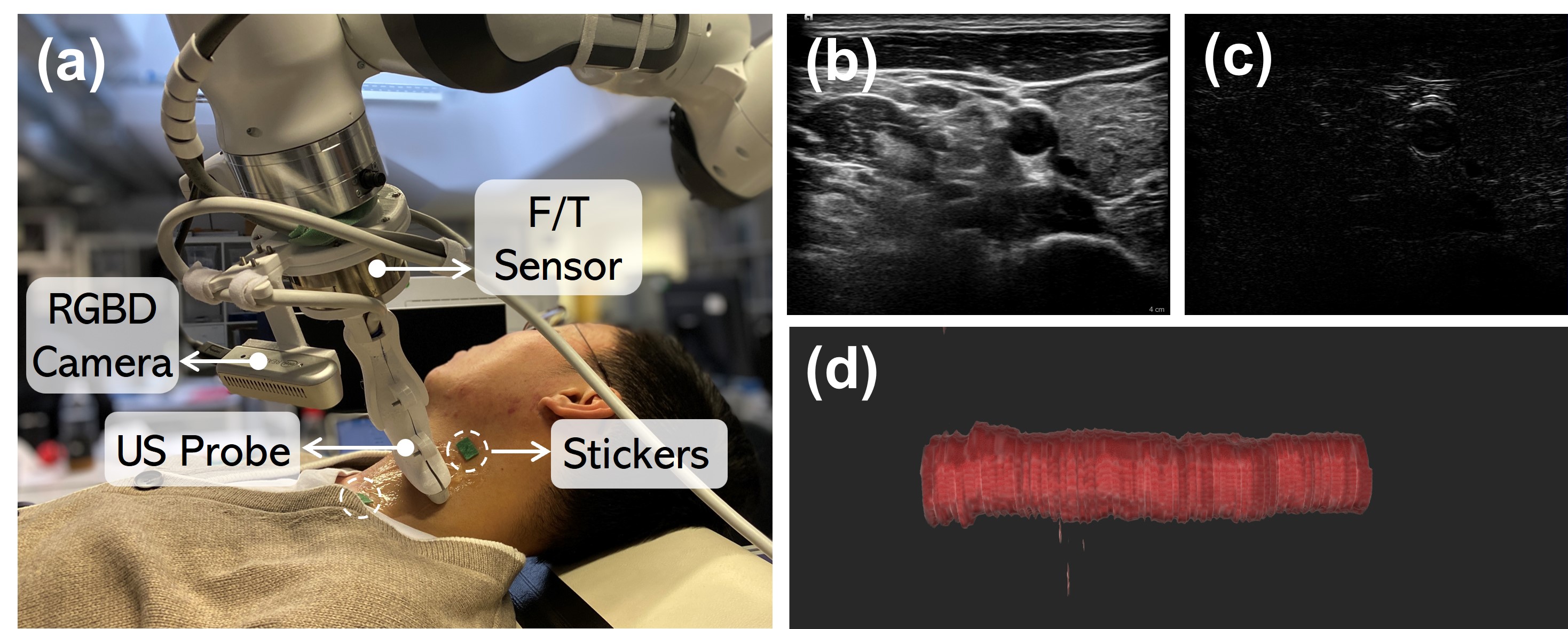}
     \caption{An illustration of the robotic carotid artery US scan scenario (\textit{Left}) and US image processing results (\textit{Right}). (a). Scan on a volunteer's neck (b). Original US image (c). Extracted pulsation map (d). 3D reconstruction of the segmentation results.}
     \label{fig:system_setup}
     \vspace{-1.0em}
\end{figure}

\par
To reliably and precisely extract the pulsation motion from a set of consecutive US images, the novel motion magnification techniques would be beneficial for enhancing such subtle motions. Wu~\emph{et al.} amplified the image intensity variations by applying spatial decomposition and temporal filtering to frames~\cite{wu2012eulerian}. Regrading medical applications, Lauridsen~\emph{et al.} extracted inapparent physiological signals (e.g., heart and respiration rate) from videography material using Eulerian video magnification (EVM)~\cite{lauridsen2019extracting}. To prevent the inadvertent damage and bleeding, Janatka \textit{et al.} explored the physiological models and applied high order video motion magnification algorithm to the laparoscopic surgery video to exaggerate the pulsation motion of the blood vessel~\cite{janatka2020surgical}. Perrot~\textit{et. al.} applied the extended EVM to the ultrafast US video to magnify the tissue motion~\cite{perrot2018video}. Similar to the scenarios where EVM was proposed, in their work, the US probe (i.e., imaging sensor) was fixed during the imaging. However, when applying EVM to the US images from a sweep scan, the movement of the US probe incurs undesired intensity variations around the tissue boundaries, which deteriorates the video magnification performance. 
To overcome this limit, a robotic US imaging pipeline is developed to acquire US images stably. Besides, inspired by video acceleration magnification \cite{zhang2017video}, we extracted the acceleration component of the US images at the temporal dimension to filter out the large linear motion.  

To the best of our knowledge, this is the first work that explicitly uses the physiological pulsation information to assist the online cross-section artery segmentation. 
Main contributions of this work are summarized below\footnote[1]{Codes for the PME and PAS-NN implementation are available at \href{https://github.com/dianyeHuang/RobPMEPASNN}{https://github.com/dianyeHuang/RobPMEPASNN}}:
\begin{itemize}
    \item [1)] In light of \cite{zhang2017video}, we proposed an intensity-based pulsation map extraction (PME) algorithm to capture real-time pulsation maps from the US image stream.
    \item [2)] We proposed a pulsation-assisted segmentation neural network (PAS-NN), which utilizes the cardiac-induced motions as guidance. It achieves better performance in the challenging task of segmenting the small radial artery in the forearm. 
    \item [3)] We present a robotic US imaging pipeline for imaging artery structures, where an external RGB-D camera is utilized to autonomously plan a scanning trajectory. 
\end{itemize}

\section{Method}
\subsection{Pulsation Map Extraction}\label{sec:pulsationMap}
In this section, instead of generating a magnified US video, we endeavour to extract the pulsation area (we termed it \textit{pulsation map}) of per US frame.
Similar to previous studies \cite{wu2012eulerian, zhang2017video}, we also assumed that:
\begin{assumption}
The image intensity variation over time corresponds to local motion.
\end{assumption}
\begin{assumption}
\label{ass:lin_motion}
Objects of no interest move approximately linearly at the temporal scale \textit{w.r.t.} the subtle motion.
\end{assumption}
\begin{assumption}
The image intensity variation over time can be approximated by a second-order Taylor series expansion:
\begin{equation}
    \mathbf{I}(\mathbf{x}+\mathbf{d}(t),~t) \approx f(\mathbf{x}) + \frac{\partial f(\mathbf{x})}{\partial\mathbf{x}}\mathbf{d}(t) + \frac{\partial^2f(\mathbf{x})}{2~\partial ^2\mathbf{x}}\mathbf{d}^2(t)
    \label{eq:Taylor_series_expand}
\end{equation}
where $\mathbf{I}(\mathbf{x},~t)$ is the spatio-temporal ultrasound image set with $\mathbf{x}=(x,~y)$ being a pixel coordinate; $\mathbf{d}(t)$ is a function representing the subtle motion over time; $f(\mathbf{x})=\mathbf{I}(\mathbf{x},~0)$ is the initial US image.In the followings, we annotate $\mathcal{L}inear=\mathbf{d}(t)\partial f(\mathbf{x})/\partial\mathbf{x}$ and $\mathcal{A}cc=0.5\mathbf{d}^2(t)\partial^2 f(\mathbf{x})/\partial ^2\mathbf{x}$ as the linear and acceleration components of the spatio-temporal ultrasound image set, respectively.
\end{assumption}

\begin{figure}[!tb]
     \centering
     \includegraphics[width=0.42\textwidth, angle=0]{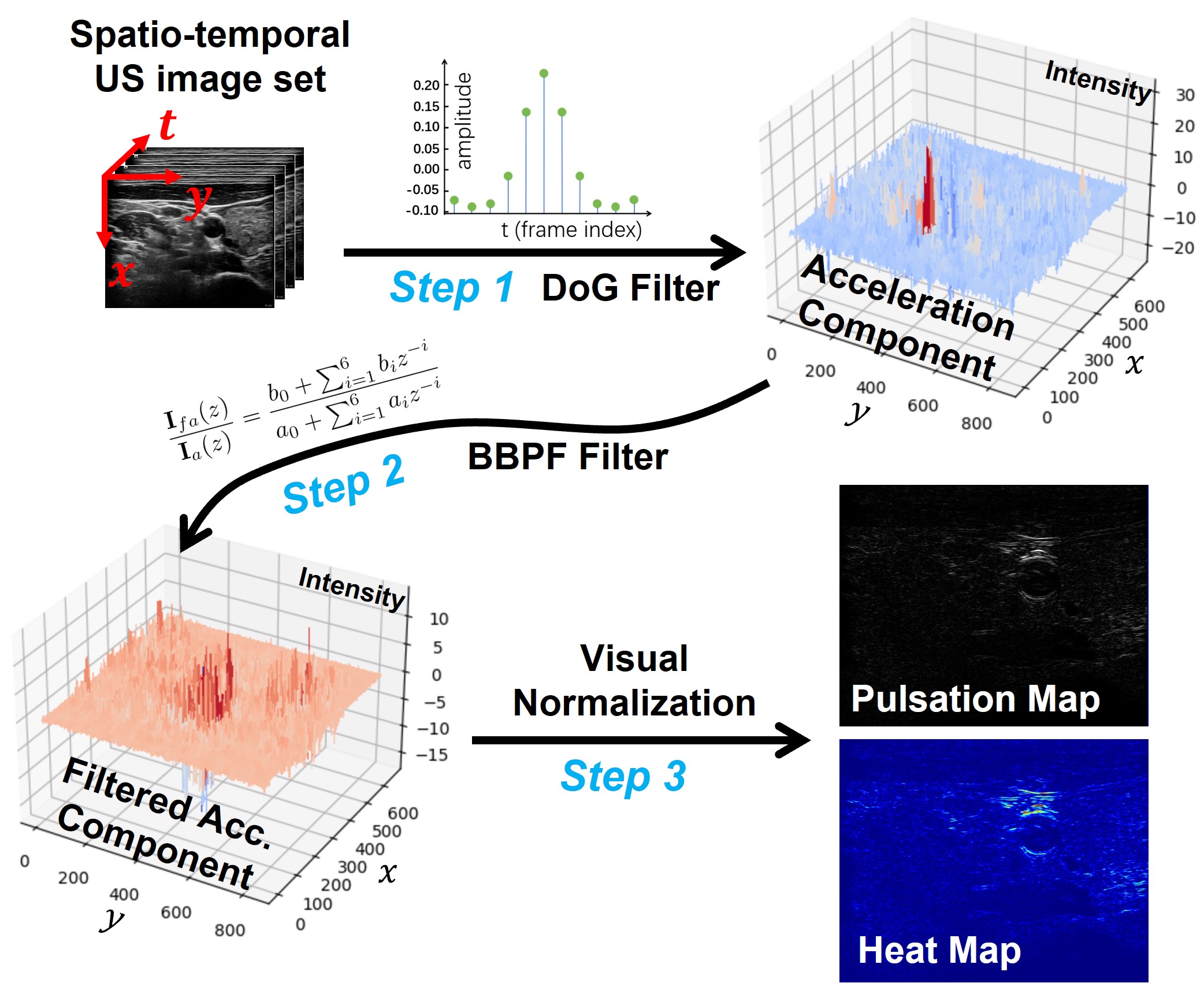}
     \caption{Steps of sequentially extracting a pulsation map from a window of
     US images (\textit{see Sec. \ref{sec:pulsationMap}}). The approximated $\mathcal{A}cc$ component in Eq. (\ref{eq:Taylor_series_expand}) and its filtered result $\mathbf{I}_{fa}$ in Eq. (\ref{eq:digital_bbpf}) are presented in a 3D space. The heat map of the pulsation map is generated by mapping the image intensity into a color bar, and it is presented here for better illustration.}
     \label{fig:pipeline_pm}
     \vspace{-0.5em}
\end{figure}

As shown in Fig. \ref{fig:pipeline_pm}
, the pulsation map is computed by the three following steps:

\subsubsection{Step 1 ($\mathcal{A}cc$ Isolation)} 
To isolate and approximate $\mathcal{A}cc$, an 1-D \textit{difference of Gaussian} ($\mathcal{D}oG$) filter is applied to the US image set along the temporal dimension as in \cite{zhang2017video}.
\begin{equation}
    \mathcal{A}cc\approx\frac{\partial^2\mathbf{I}(\mathbf{x},~t)}{\partial t^2\otimes \mathcal{G}(\sigma,~t)} = \mathbf{I}(\mathbf{x},~t)\otimes \frac{\partial^2\mathcal{G}(\sigma,~t)}{\partial t^2} 
    \label{eq:hh}
\end{equation}
where $\otimes$ is a convolution operator and $\mathcal{G}(\sigma, ~t)$ denotes a Gaussian filter with variance $\sigma^2$. $\mathcal{D}oG=\mathcal{G}(\sigma/2,~t)-\mathcal{G}(2\sigma,~t)$ involves the subtraction of a blurred signal from a less blurred one, which can be considered as an approximation to the Laplacian of Gaussian filter. Thus, we employed $\partial^2\mathcal{G}(\sigma,~t)/\partial t^2\approx\mathcal{D}oG$ to reduce the computation burden, where $\sigma=f_s/(4f_d\sqrt{2})$, $f_s$ and $f_d$ represent the sampling rate and interested motion frequency, respectively. 
\begin{figure*}[!tb]
     \centering
     \includegraphics[width=0.90\textwidth, angle=0]{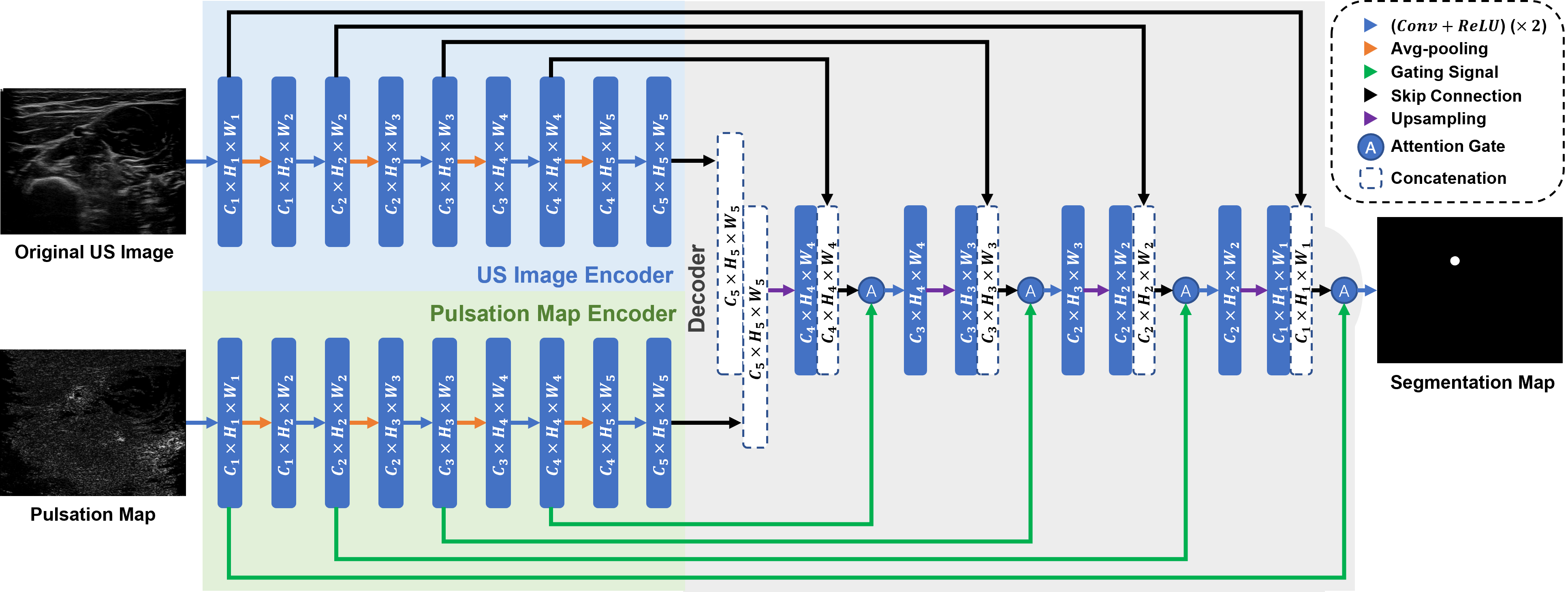}
     \caption{An illustration of the proposed PAS-NN structure with the original US image and the proposed pulsation map being the input signals. The computation of the pulsation map and the inference of the segmentation result are performed in real time accelerated by a GPU.}
     \label{fig:pms_nn}
\end{figure*}
\subsubsection{Step 2 (Bandpass Filtering)} Butterworth bandpass filter (BBPF) further refines the extraction result by constraining the output from \textit{Step 1} to the desired frequency band. As the normal pulsation rate for healthy adults ranges from $60$ to $100~bpm$ (\textit{i.e.} $1.0\sim 1.7~Hz$), we select $f_d=1.5~Hz$, $f_s=30.0~Hz$ in \textit{Step 1} and a third-order digitized BBPF (\textit{see Eq. (\ref{eq:dtf_bbpf})}) with a passband of $0.9\sim2.0~Hz$ and a sampling rate of $30.0~Hz$:
\begin{equation}
    \frac{\mathbf{I}_{fa}(z)}{\mathbf{I}_a(z)}=\frac{b_0 + \sum^{6}_{i=1}b_i z^{-i}}{a_0 + \sum^{6}_{i=1}a_i z^{-i}}
    \label{eq:dtf_bbpf}
\end{equation}
Finally, the filtering process is realized by Eq. (\ref{eq:digital_bbpf}):
\begin{equation}
    \mathbf{I}_{fa}^{(n)}(\mathbf{x})=\sum_{i=0}^6\left(\frac{b_i}{a_0}\mathbf{I}_{a}^{(n-i)}(\mathbf{x})\right)-
    \sum_{i=1}^6\left(\frac{a_i}{a_0}\mathbf{I}_{fa}^{(n-i)}(\mathbf{x})\right)
    \label{eq:digital_bbpf}
\end{equation}
where $\mathbf{I}_{a}^{(k)}(\mathbf{x})$ and $\mathbf{I}_{fa}^{(k)}(\mathbf{x})$ denote the $k$-th input image frame and its filtered result; $a_i$, $b_i$ are coefficients of Eq. (\ref{eq:dtf_bbpf}).

\subsubsection{Step 3 (Visual Normalization)} To convert the filtered $\mathcal{A}cc$ (\textit{i.e.} $\mathbf{I}_{fa}$) into a visualizable image, the following transformations are performed:
\begin{equation}
    \mathbf{I}_{pm}(\mathbf{x}) = \mathcal{L}im\left(\mathcal{G}amma\left(\alpha\Big|\mathbf{I}_{fa}(\mathbf{x})\Big|,~\gamma\right), ~0,~255\right)
    \label{eq:vis_pm}
\end{equation}
where $\mathbf{I}_{pm}(\mathbf{x})$ represents the pulsation map, $\mathcal{L}im(i,~0,~255)=\max(\min(i,~255),~0)$ constrains the pixel intensity, $\mathcal{G}amma(i,~\gamma)=i^{1/\gamma}$ denotes the gamma transformation with $\gamma\in(0.0,~1.0)$ exaggerating the intensity difference for better visualization, $\alpha~(>1.0)$ is a magnification coefficient. The choices of magnification and gamma coefficients vary with the acquisition settings of the US machine and the scan target tissue. In this work, we empirically chose $\alpha=38$, $\gamma=0.80$ for the carotid artery scan, and $\alpha=20$, $\gamma=0.65$ for the radial artery scan.


\subsection{Pulsation-Assisted Segmentation Network}
The periodic cardiac-induced motion helps to locate the arteries since it is hard to tell whether a vessel is a vein or an artery from one static US image, especially for small arteries such as the radial artery. The proposed pulsation map, as defined in \textit{Sec.~\ref{sec:pulsationMap}}, highlights the regions where the cardiac-induced pumping motion appears, which roughly represents the positions of arteries in the US image. In light of the fact that the periodic motions introduced by heartbeat are taken as guidance by the human operators when localizing the positions of arteries in US images, we proposed pulsation-assisted segmentation neural network (PAS-NN). 

As shown in Fig \ref{fig:pms_nn}, two separate encoders are implemented to extract the meaningful features from the original US image and its corresponding pulsation map. The bottlenecks of the two encoders are concatenated and fed into a decoder to generate the segmentation maps. Attention gate models~\cite{schlemper2019attention} are implemented at each scale. The upsampled features are concatenated with the skip connections \cite{ronneberger2015u} from the image encoder, while the extracted features from the pulsation map are handled as attention gating signals in different scales to provide guidance to the network. 

\begin{figure}[!tbp]
     \centering
     \includegraphics[width=0.45\textwidth, angle=0]{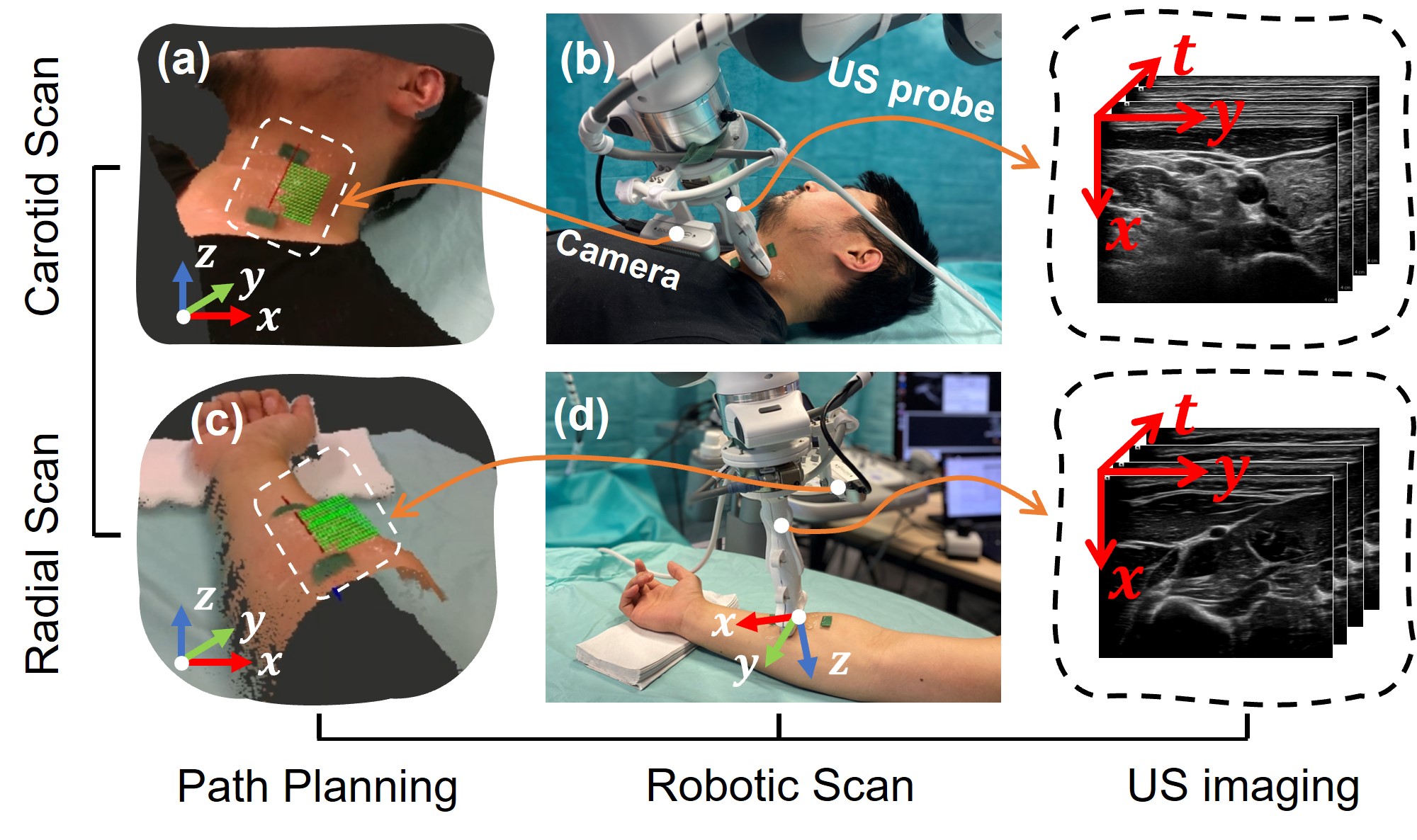}
     \vspace{-1.0em}
     \caption{An illustration of robotic ultrasound imaging pipeline for carotid and radial artery scans. 
     During the scan, a hybrid force motion controller is activated to simultaneously regulate the contact force measured along the probe-axis and the scan velocity. The US images are streamed alongside the pose of the US probe's coordinate frame (x-axis: normal to the imaging plane; z-axis: aligned with the US probe-axis).}
     \label{fig:pipeline_imaging}
     \vspace{-0.8em}
\end{figure}


\begin{table*}
\centering
\caption{Performance comparison of the proposed PAS-NN with three other\\
    segmentation networks on the US carotid artery dataset and radial artery dataset.}\label{tab1}
  \resizebox{\textwidth}{!}{
      \begin{tabular}{L{0.09\textwidth}C{0.09\textwidth}C{0.09\textwidth}C{0.09\textwidth}C{0.09\textwidth}|C{0.09\textwidth} C{0.09\textwidth} C{0.09\textwidth} C{0.09\textwidth}}
        \toprule
        \multirow{2}{*}{Method} &
          \multicolumn{4}{c|}{Carotid} &
          \multicolumn{4}{c}{Radial}\\
          \cmidrule{2-9}
          & {Dice} & {Precision} & {Recall} & {Hausdorff} & {Dice} & {Precision} & {Recall} & {Hausdorff} \\
          \midrule
        UNet~\cite{ronneberger2015u}            & $\textbf{0.954}\pm 0.012$ & $\textbf{0.942}\pm 0.021$ & $0.968\pm 0.023$ & $1.772\pm 0.278$ & $0.780\pm 0.155$ & $0.863\pm 0.143$ & $0.729\pm 0.176$ & $2.045\pm 0.611$\\
        GLFRNet~\cite{song2022global}           & $0.941\pm 0.017$ & $0.933\pm 0.032$ & $\textbf{0.980}\pm 0.016$ & $\textbf{1.749}\pm 0.344$ & $0.591\pm 0.160$ & $0.539\pm 0.201$ & $0.734\pm 0.177$ & $3.108\pm 0.702$\\
        Att-UNet~\cite{schlemper2019attention}  & $\textbf{0.954}\pm 0.012$ & $\textbf{0.942}\pm 0.021$ & $0.968\pm 0.023$ & $1.772\pm 0.278$ & $0.655\pm 0.350$ & $0.794\pm 0.342$ & $0.595\pm 0.339$ & $2.261\pm 0.854$\\
        PAS-NN                 & $\textbf{0.954}\pm 0.011$ & $0.940\pm 0.022$ & $0.976\pm 0.016$ & $1.755\pm 0.232$ & $\textbf{0.892}\pm 0.106$ & $\textbf{0.941}\pm 0.089$ & $\textbf{0.859}\pm 0.125$ & $\textbf{1.633}\pm 0.525$\\
        \bottomrule
      \end{tabular}
  }
  \vspace{-0.5em}
\end{table*}

\subsection{Robotic Ultrasound Imaging}
\subsubsection{Experimental Setup}
As depicted in Fig. \ref{fig:system_setup} (a), a redundant robotic arm (Franka Emika Panda, Franka GmbH) equipped with an RGB-D camera (Intel$^\circledR$ Realsense\texttrademark ~D435), force-torque (F/T) sensor (GAMMA, SI-32-2.5/SI-65-5, ATI) and a linear US probe (ACUSON Juniper, SIEMENS AG) is utilized as a platform to acquire US images via a frame grabber (MAGEWELL). The robotic arm is controlled by a laptop (AMD Ryzen 9 5900HX CPU, NVIDIA GeForce RTX 3070) with the robotic operating system (ROS) running on the Ubuntu 20.04 system.

\subsubsection{Robotic Scan}
Fig. \ref{fig:pipeline_imaging} presents the carotid (\textit{first row}) and radial (\textit{second row}) artery scanning scenarios. The scanning positions and orientation of the US probe are determined by processing the point cloud captured by the RGBD camera.
Initially, the camera will detect two green stickers that are manually attached to the volunteer's skin surface. These two points on the point cloud are then linearly interpolated to determine the positions for scanning. We make the assumption that the scanning surface can be treated as a rectangular plane. Consequently, the orientation of the US probe remains constant during the scan. The z-axis is determined by calculating the average of the estimated normal vectors perpendicular to the skin surface at the scanning positions. The x-axis is determined by the positions of the two stickers.
During the scan, a hybrid force position controller is activated to ensure a consistent contact force. In this work, the desired contact forces for carotid and radial artery scanning are $3.0~N$ and $5.0~N$, respectively. They are empirically set to guarantee the acoustic coupling while minimizing tissue deformation. To fulfill the \textit{Assumptions} described in \textit{Sec.} \ref{sec:pulsationMap} to the greatest extent, the scan velocity is set at $0.5~mm/s$. 
\begin{figure}[!t]
     \centering
     \includegraphics[width=0.45\textwidth, angle=0]{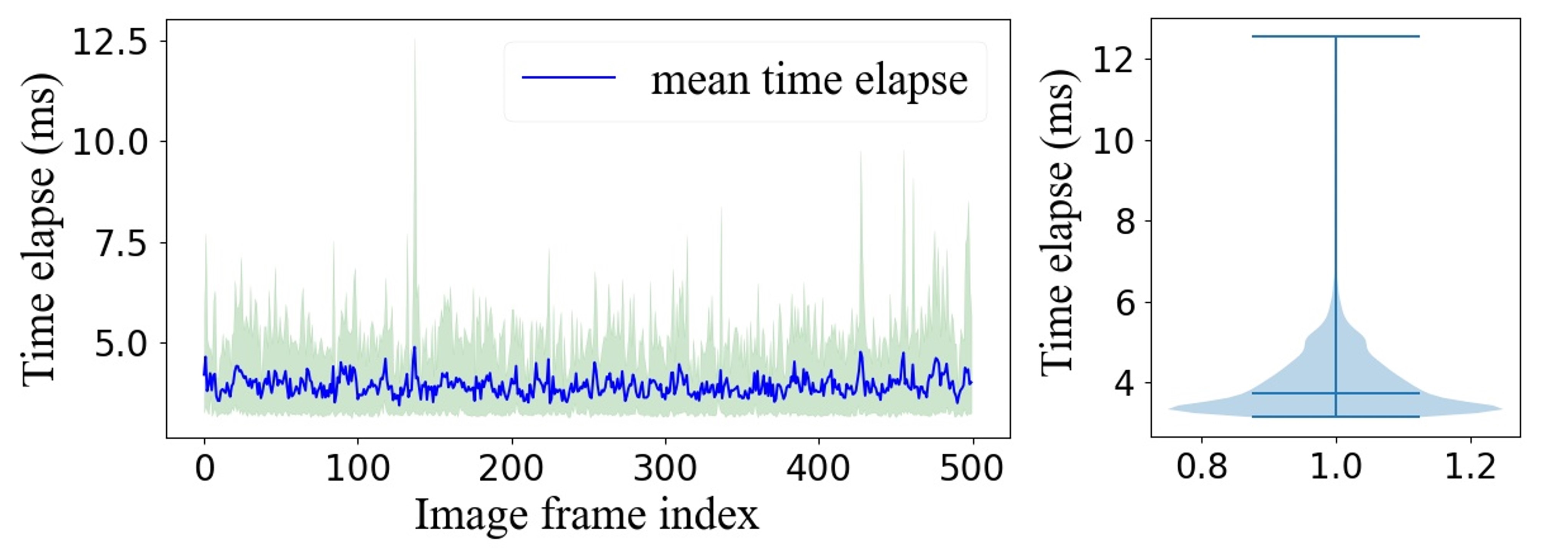}
     \vspace{-0.5em}
     \caption{Time elapses of computing a pulsation map. \textit{Left:} the green shadow indicates the minimal and maximal time elapses at the distinct image frame; \textit{Right:} a violin plot showing the time elapses distribution.}
     \vspace{-1.0em}
     \label{fig:time_ellapse}
\end{figure}
\subsubsection{Artery Segmentation}
The US images and the poses of the US probe's tip were recorded at $30.0~Hz$ and $70.0~Hz$, respectively. These signals were synchronized based on the timestamps.   
Given that the pulsation map is a new modality incorporated into the segmentation pipeline, it is crucial to know the computing time, irrespective of the segmentation network architecture. The computation of pulsation map is accelerated by the \textit{NVIDIA GeForce RTX 3070 GPU} with the help of the \textit{pyTorch} package. The Time elapsed for computing a pulsation map is presented in Fig. \ref{fig:time_ellapse}, which shows the statistic result from the continuous processing of $9$ groups of US image sets. Each group consists of $500$ frames with a resolution of \textit{$657\times837$}. The violin plot shows that the time consumption mainly distributes under $7.0~ms$ per image. 
During the inference of PAS-NN, the computing of the pulsation map and the segmentation of the artery are performed in real-time, accelerated by the \textit{NVIDIA GeForce RTX 3070 GPU}. 

\section{Experiments and results}
\begin{figure*}[!tb]
     \centering
     \includegraphics[width=0.9\textwidth, angle=0]{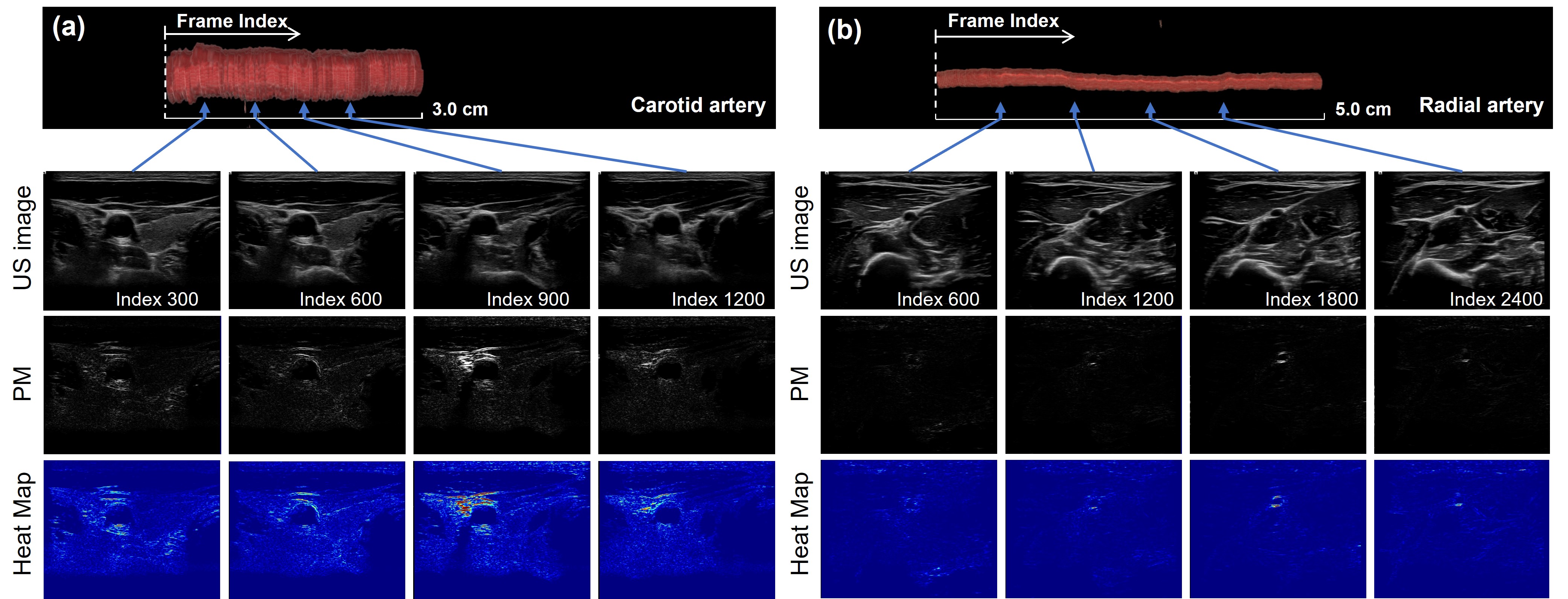}
     \caption{3D reconstruction of the segmentation results from the PAS-NN for (a). the carotid artery and (b). radial artery. From the second to the last rows present the original US image, the proposed pulsation map (PM), and the heat map of PM, respectively. From the left to the right columns show selected snaps of the intermediate results. Heat maps are presented here for better visualization. The 3D reconstruction process was performed offline using the \textit{ImFusion Suite} (ImFusion GmbH, Munich, Germany).}
     \label{fig:pm_res}
     \vspace{-0.5em}
\end{figure*}
\subsection{Results of the PAS-NN Segmentation}
The US data of carotid and radial arteries is acquired from three adult volunteers aged $29\sim32$ using two default sets of US parameters provided by the US machine for carotid and radial artery acquisition, respectively\footnote[2]{The acquisition was performed within the Institutional Review Board Approval by the Ethical Commission of the Technical University of Munich (reference number 244/19 S).}. When extracting the pulsation map, parameters in Eq. (\ref{eq:vis_pm}) are selected empirically where $\alpha=38.0$, $\gamma=0.80$ for carotid and $\gamma=0.65$ for radial. The annotation is carefully done under the supervision of our clinical partners. The training data includes the images from two of the three volunteers, and one is left for testing.
The training datasets for carotid and radial arteries have the same dataset size of $3000$, while the testset size is set to $800$. For each network structure, two networks are trained for carotid and radial artery segmentation, accordingly. All the networks are trained using Adam optimizer with a fixed learning rate of $1\times10^{-4}$ for $100$ epochs. 

Table~\ref{tab1} shows the comparison results between the proposed PAS-NN and other network structures. All the networks perform equally well for the carotid segmentation task by achieving more than $0.94$ Dice scores. Such good performance can be expected because the carotid artery's cross-section is large, and there are barely two similar anatomic shapes of this size in the same US image. This fact indicates that the carotid artery can be easily detected from the US images without further information. However, in the case of radial artery segmentation, the proposed PAS-NN outperforms all the other networks with the highest scores. In our dataset, the average diameter of the radial artery is $2.51\pm 0.16~mm$. Compared to the carotid artery, the radial artery is much smaller and less apparent in US images.  These observations demonstrate that when the size of the vessel is small and hard to be directly identified from the US image, the involvement of a pulsation map can provide essential guidance to the network to locate the relevant regions and thus increases the robustness of the network.

\begin{figure}[!t]
     \centering
     \includegraphics[width=0.45\textwidth, angle=0]{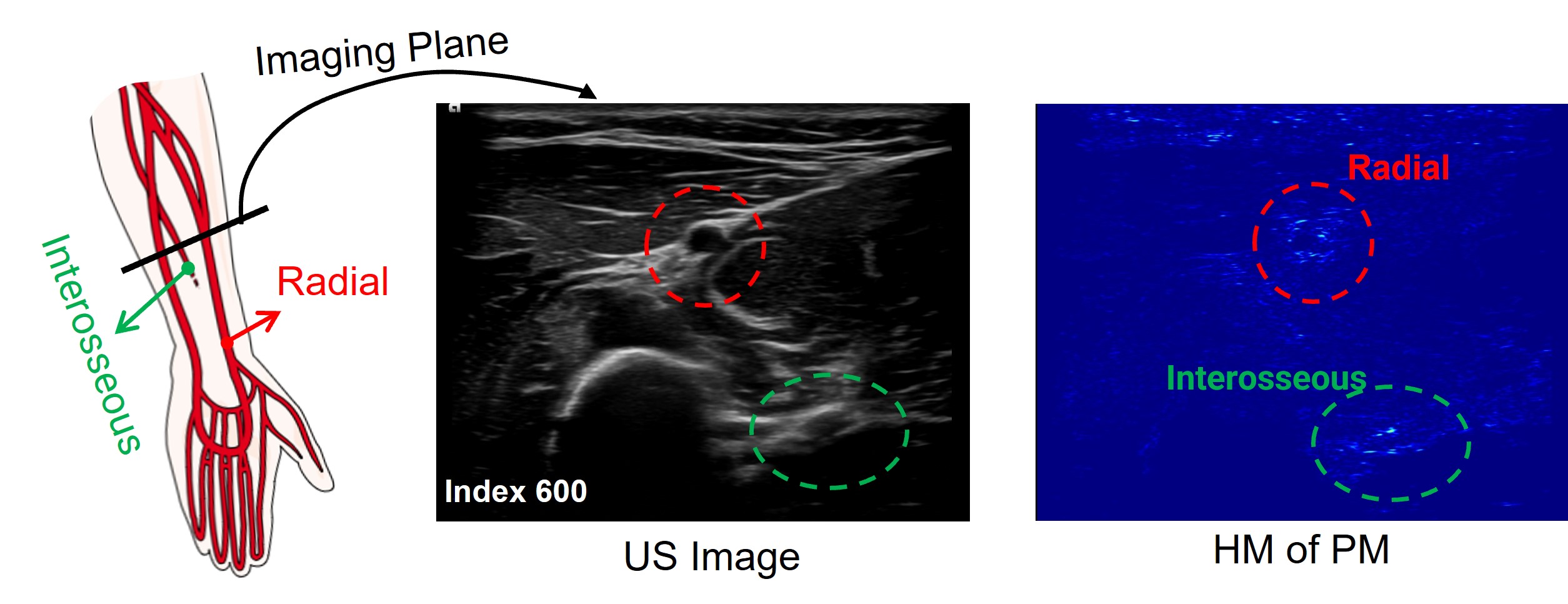}
     \vspace{-1.0em}
     \caption{An illustration of the pulsation map, PM (\textit{Right}) of the radial artery ultrasound image (\textit{Middle}) acquired at the described imaging plane (\textit{Left}).}
     \label{fig:vis_arteries}
     \vspace{-0.8em}
\end{figure}

\subsection{Overall Results of PME}
After the PAS-NN is well trained, as depicted in Fig. \ref{fig:pipeline_imaging}, we performed robotic ultrasound scans of a volunteer's carotid artery and radial artery, respectively. Fig. \ref{fig:pm_res} presents the 3D reconstruction results of the arteries (\textit{see the top row}) with selected snaps of the intermediate results, including the original US image, its corresponding pulsation map and heat map. The heat map is identical to the pulsation map but with a different color mapping for better visualisation.

Although in both cases, the pulsation motions are extracted successfully, the pulsation of the carotid artery is more visually apparent than the radial artery in the pulsation map. This result is aligned with the fact that compared with the radial artery, the carotid artery occupies a larger area of the US image, thus raising a wider range of image intensity changes. 

From the heat maps of the pulsation map (\textit{see the third row of Fig. \ref{fig:pm_res}}), we can find that the artery lumen is highlighted. Regardless of the intensity changes induced by the movement of the US probe, this highlighting is caused by the pulsation of the artery which leads to periodic changes of the image intensity around the artery lumen locations. Such intensity changes are captured and amplified by the proposed PME algorithm. However, it is only partially highlighted because of two reasons. From the carotid pulsation maps, we can see that the boundary of the carotid artery is not always well imaged (i.e., not all the boundary is clearly captured by the US probe) due to the imaging pose of the probe, contact forces between the probe and human skin or acquisition parameter settings of the US machine etc.. Besides, for distinct imaging poses, partial boundary of the artery (e.g., the vertical boundary of the radial artery in Fig. \ref{fig:pm_res} (b) indexed $1800$) does not have an apparent movement (\textit{i.e.,} the surrounding image intensity has little variations) as time evolves. 

We can also find that both the target artery lumen and the surrounding tissue borders are amplified (\textit{see Fig. \ref{fig:pm_res} (a) pulsation map of US image indexed $300$}). Two reasons can explain this observation; one reason is that the movement of the US probe happens to induce temporal variations of image intensity lying in the setting passband, and the other reason is that the pulsation of the artery raises the movement of its surrounding tissues. The former reason will be discussed in \textit{Sec.~\ref{subsec:lin_res}} while the evidence for the latter reason can be shown by comparing the pulsation map of the US images in Fig. \ref{fig:pm_res} (a) indexed $900$ and in Fig. \ref{fig:pm_res} (b) indexed $1800$, where the pulsation of radial artery is less salient and therefore does not cause much movement of the adjacent tissue borders.

To verify the validity of the proposed PME algorithm, we further examined the pulsing regions highlighted by the computed pulsation map. We found two flashing areas in the pulsation maps of the radial artery scan (\textit{see the red and green circles pointed out in Fig. \ref{fig:vis_arteries}}). Besides the target radial artery, an interosseous artery (artery between the bones) is also pulsing alongside at the right bottom of the acquired US image, which can hardly be noticed. This interpretation further proves the effectiveness of the proposed PME algorithm. However, it is notable that this result, in most cases, could be obtained only when the artery anatomy is well captured in a window of sequential US images, and the US probe should be kept as still as possible. Under these circumstances, one could consider taking advantage of the PME algorithm to locate unapparent arteries during US imaging. 

\subsection{Comparison between the PME and Linear Methods}
\label{subsec:lin_res}
\begin{figure}[!t]
     \centering
     \includegraphics[width=0.45\textwidth, angle=0]{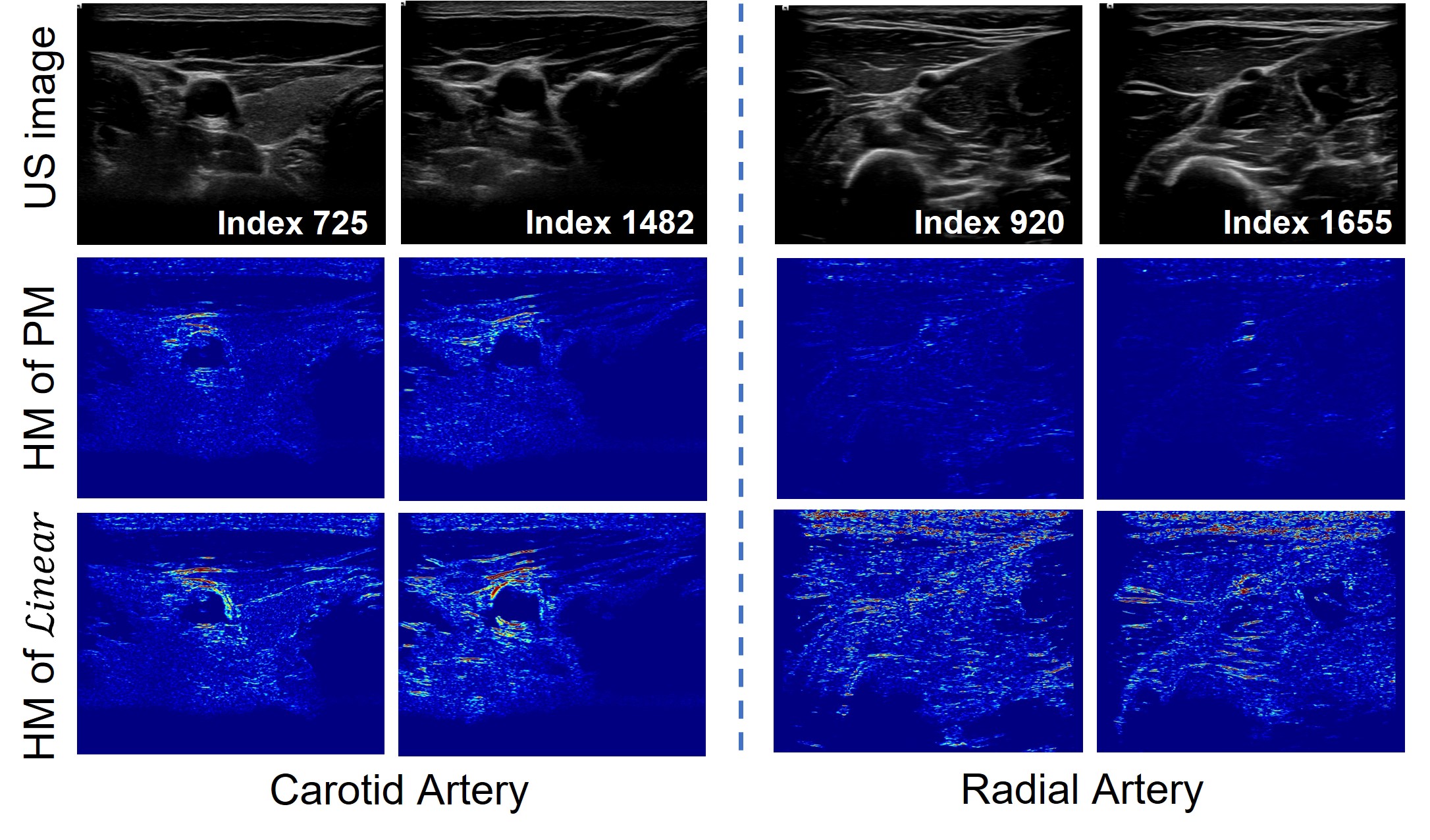}
     \vspace{-1.0em}
     \caption{Comparison between the extracting results of the $\mathcal{A}cc$ and $\mathcal{L}inear$ components in Eq. (\ref{eq:Taylor_series_expand}). The second and third rows represent the heat map (HM) of the acceleration (proposed pulsation map, PM) and the linear results, respectively.}
     \label{fig:acclin_cmp}
     \vspace{-0.5em}
\end{figure}

To demonstrate the advantage of extracting the acceleration part of the intensity variations, we compared the results of extracting the pulsation maps of the linear $\mathcal{L}inear$ and acceleration $\mathcal{A}cc$ components as formulated in Eq. (\ref{eq:Taylor_series_expand}). As shown in Fig. \ref{fig:acclin_cmp}, the heat map of the $\mathcal{L}inear$ component indicates more irrelevant pulsation regions than the $\mathcal{A}cc$ (proposed) component. The proposed pulsation map utilizes the acceleration nature of eliminating the constant linear motions, the manner in which the irrelevant movements are assumed to follow (\textit{see Assumption \ref{ass:lin_motion} in Sec.~\ref{sec:pulsationMap}}). Therefore, the PME algorithm can reduce the noises of the pulsation map to some extent.

\section{Discussion}
In this work, we implement the proposed PME algorithm to extract the pulsation regions of US images automatically streamed by a redundant robotic arm. Unlike the Perrot \textit{et. al.}'s work \cite{perrot2018video}, which first applied a video magnification algorithm to ultrafast US images; in our work, the US images are streamed at a relatively low sampling frequency $30.0~Hz$. Instead of keeping the imaging sensor at the same position, the US probe moves continuously at a constant velocity. Furthermore, the real-time computation of the pulsation map is also expected to fit the pulsation signals into various potential online applications. 

From the result of this work, we also find some distinct differences between amplifying the optical video (\textit{e.g.,} laparoscopic surgical videos in \cite{janatka2020surgical}) and the US sweep. Streaming US images involves contact interaction, imaging plane selection, and the choice of US acquisition parameters for the US machine, which highly affects the image quality. Noted that the pulsation of the target anatomy (carotid artery especially) distorts the surrounding tissue and, therefore, raises another challenge of differentiating between the active motion of the target and the passive motion of the tissues.

\section{Conclusion and Future Work}
This work presents an autonomous robotic artery US imaging system that can perform online artery segmentation enhanced by the extracted pulsation information. The proposed PME algorithm extracts the pulsation signals, and the computation time is less than $7~ms$ per frame on average. US scans on a volunteer's neck and forearm are performed, and the carotid and radial artery are segmented online. The results demonstrate that the proposed PAS-NN achieve comparable performance on the carotid artery segmentation and outperforms other state-of-the-art networks in the task of radial artery segmentation. 

Utilizing a pulsation map as an extra input signal for a segmentation network is only one of its use cases. As the pulsation map is generated in real-time, it can also serve as a feedback signal for the robotic artery tracking application or become an artery detector when the segmentation network fails to track the target. Thanks to the dynamic imaging property of the US imaging modality, the pulsation map can also be applied to observe other anatomy guided by the human body's physiological motions, including cardiac-induced pulsation (our case), respiration, and bowel movements. Extracting these motion patterns can assist robotics in interpreting the acquired images and take a more reasonable action to complete the corresponding tasks.


\bibliographystyle{IEEEtran}
\bibliography{IEEEabrv, ref}

\end{document}